\documentclass[amsmath,amssymb,amsfonts,aps,pre,preprint,superscriptaddress,bibnotes,showpacs,showkeys,longbibliography,nofootinbib]{revtex4-1}
\usepackage{epsfig}
\usepackage[english]{babel}
\usepackage{color}

\newcommand{\sgn}{\operatorname{sgn}}

\begin{document}
\title{Engines with ideal efficiency and nonzero power\\ for sublinear transport laws}

\author{Jesper Koning}
\affiliation{Instituut voor Theoretische Fysica, Celestijnenlaan 200D, KU Leuven, B-3001 Leuven, Belgium}

\author{Joseph O. Indekeu}
\affiliation{Instituut voor Theoretische Fysica, Celestijnenlaan 200D, KU Leuven, B-3001 Leuven, Belgium}

\date{\today}
\begin{abstract}
It is  known that an engine with ideal efficiency ($\eta =1$ for a chemical engine and $e = e_{\rm Carnot}$ for a thermal one) has zero power because a reversible cycle takes an infinite time. However, at least from a theoretical point of view, it is possible to conceive (irreversible) engines with nonzero power that can reach ideal efficiency. Here this is achieved by replacing the usual linear transport law by a sublinear one and taking the step-function limit for the particle current (chemical engine) or heat current (thermal engine) versus the applied force. It is shown that in taking this limit exact thermodynamic inequalities relating the currents to the entropy production are not violated.
\end{abstract}
\maketitle

\section{Context}
The question whether or not it is possible to construct a thermodynamic engine that can retain nonzero power while operating at maximal (ideal) efficiency, is a subject of much recent debate. This question is most often posed in the framework of thermal engines, in which one asks whether the Carnot limit is feasible at finite power. For classical thermal engines a proof has been formulated which implies that the entropy production must vanish in the Carnot limit \cite{S}. This theorem rules out finite power in the Carnot limit, provided the particle dynamics is regular with a finite damping constant. Also for classical thermal engines it has been shown that in microscopic systems subject to random fluctuations it is possible to operate very close to Carnot efficiency in a transient state \cite{P}. This is possible (only) in a singular coupling limit, physically corresponding to a working fluid with divergent current correlations, reminiscent of bulk criticality. In this state the familiar linear current-force relation is inadequate. 

The idea that bulk criticality might play a crucial role in attempting to maintain finite power in the Carnot limit, has been developed more thoroughly in the context of quantum thermal engines \cite{CF,M}. An array of $N$ quantum Otto cycles endowed with a divergent specific heat and with critical speeding up of the relaxation time, may approach the Carnot limit for $N \rightarrow \infty$, while the power $P$ per device, $P/N$ is kept constant \cite{CF}. Also noteworthy is that a critical nanoscale quantum Schr{\"o}dinger junction can reach efficiencies at maximum power that exceed the (presumably exact) bounds valid for similar but classical systems, in and beyond linear response \cite{M}.   

Still in the context of quantum thermal engines, other mechanisms have been uncovered that may boost the power and/or the efficiency. In \cite{A} a quantum many-body system is considered with $N$ particles. It is found that, if the system-bath Hamiltonian is especially chosen to depend on the system Hamiltonian, and if also the bath Hamiltonian is tailored, then a finite power per particle can be achieved. Under these circumstances Carnot efficiency can be reached asymptotically, for $N \rightarrow \infty$. Another mechanism has been proposed in \cite{LA}, where it is shown how to enhance the performance of a quantum Otto cycle by employing the two thermal reservoirs to produce an out-of-equilibrium electromagnetic field. The quantum Otto engine's efficiency is at most the Carnot efficiency, but the enhanced version has an efficiency that can even reach unity. Furthermore, for the usual cycle the power vanishes at Carnot efficiency, but in the enhanced design finite power (but not maximal power) is possible even at unitary efficiency. The efficiency at maximal power does not reach unity (but peaks at about 0.8). 

Within the framework of (isothermal) chemical engines, the possibility of achieving ideal (in this case unitary) efficiency at finite power  
has also been pursued recently. A kinetic network relevant to chemical motors or biochemical processes 
has been conceived characterized by a vanishing entropy production while exhibiting a finite dissipative current \cite{EP}. This is possible when the current occurs on a much longer time scale than the driving by external forces (through reversible pumps). In a limiting case unitary efficiency at maximum (and finite) power can be reached.  

Our present contribution in this context is concerned with macroscopic classical systems (in the thermodynamic limit) and is twofold. Firstly, we complement previous work devoted to chemical engines with an analytical calculation, pointing out properties that were hitherto not noticed, and secondly we show by analytical calculations and numerical computations that these properties have analogues for thermal engines.

Ideal efficiency is a well-known concept for thermal and chemical engines. For thermal engines it is the efficiency $e_c \equiv e_{\rm Carnot} = 1- T_{\ell}/T_h$ obtained  for a reversible Carnot cycle operating between absolute temperatures $T_h$ and $T_{\ell}$ ($<T_h)$. For simple chemical engines, involving particle exchange but no chemical reactions\footnote{For concreteness, an elementary example of the type of chemical cycle we have in mind would be a balloon attached to the bottom of a bucket partly filled with water at constant temperature. Particles are taken up by the balloon by connecting its valve to pressurized air. This causes the water level to rise and enables the ``engine" to perform mechanical work. Particles are  subsequently released by opening the valve in ambient air, until the balloon returns to its initial volume. And then again.}, it is the efficiency $\eta = 1$ obtained for a reversible isothermal cycle operating between chemical potentials $\mu_h$ and $\mu_{\ell} $ ($< \mu_h)$. These engines, however, have zero power because a reversible cycle takes an infinitely long time. Therefore, by including irreversible finite-time processes in these cycles, while keeping them simple, one can discuss how to maximize their power and compute their {\em efficiency at maximum power}. This has led to remarkable and universal results for thermal \cite{CA} and chemical \cite{Chinese,longpaper} engines.

There are noteworthy conceptual differences between the two types of cycles. In the thermal cycle some of the heat that is taken up at $T_h$ must be released at $T_{\ell}$ and cannot be used to do useful work $W$, in accord with the second law of thermodynamics. In order to have $W \neq 0$ one must have $T_h \neq T_{\ell}$. However, for the isothermal chemical engine the chemical energy of particle exchange can be entirely converted into useful work $W$. Some heat is taken up, but the same amount of heat is released, at one and the same temperature $T$. Notwithstanding the isothermal setting, the cycle displays a nonzero area in a $PV$-diagram, and therefore $W \neq 0 $. Note that uptake and release of particles is isobaric {\em and} isothermal \cite{longpaper}. 

The  irreversible finite-time processes that are included to yield cycles with nonzero power, are heat transport by conduction \cite{CA} for the thermal cycle and particle diffusion or effusion \cite{Chinese,longpaper} for the chemical one. These transport processes involve new, intermediate, temperatures or chemical potentials with respect to which the power of the cycle can be maximized. The main (universal) result is that the efficiency at maximum power  equals half the ideal efficiency \cite{CVDB}. This is an explicit illustration of the intuitive insight that, in order to achieve nonzero power, one must sacrifice efficiency.

\section{Main result for the chemical engine}
There is an interesting property emerging in these recent works and we point out that new physics arises in conjunction with it.  This property is concerned with the fact that the efficiency at maximal power can be higher than half the ideal efficiency, and can even approach ideal efficiency, when the transport process is modeled by a nonlinear law, in particular a sublinear one. 

A nonlinear transport law should not be viewed as just an academic exercise. It can be physically relevant. It has recently become clear that linear transport coefficients may {\em diverge} under certain circumstances. The thermal conductivity of a fluid, for example, diverges at the bulk critical point \cite{Sengers1985,Sengers2014,Anisimov}. Therefore, a description in terms of linear response becomes physically unacceptable and, in this case, a sublinear counterpart can be meaningfully considered.  

The nonlinear law that has been proposed in this context \cite{longpaper}, is characterized by an exponent $\theta \geq 0$ in the generalized transport equation
\begin{equation}\label{chemtrans}
\frac{dN}{dt}=\lambda_i k_BT\sgn(\mu_i-\mu_i^*) \left |\frac{\mu_i-\mu_i^*}{k_BT}\right |^\theta,
\end{equation} 
for the particle current in a chemical engine. Here, $N$ is the number of particles, $\lambda_i$ is a transport coefficient ($i=h$ for uptake and $i=\ell$ for release of particles) and $k_B$ is Boltzmann's constant. Note that $\theta = 1$ corresponds to usual linear transport as described by Fick's law. A sublinear law is obtained for $\theta < 1$. The chemical potentials $\mu_i$ pertain to the reservoirs and the $\mu_i^*$, with $\mu_\ell \leq \mu_\ell^* \leq  \mu_h^* \leq \mu_h$, are adjustable intermediate values.
For this transport law, the efficiency at maximal power satisfies $\eta_{mp} =1/(1+\theta)$ \cite{longpaper}, which approaches unity in the limit of a step-function transport law. The new physics we now link to this property is that ideal efficiency $\eta =1$ can be achieved without giving in on the power of the engine. This is different from the intuitively reasonable guess that the power might go to zero when the efficiency resumes its ideal value \cite{JN}. Surprisingly, the maximum power as a function of $\theta$ is itself maximal at $\theta = 0$. 

\section{Main result for the thermal engine}
For the thermal engine, the generalized transport law we propose is
\begin{equation}\label{heatmodified}
\frac{dQ}{dt}=\kappa_i \bar T\sgn(T_i-T_i^*) \left |\frac{T_i-T_i^*}{\bar T}\right |^\theta,
\end{equation} 
for the heat current. Here, $Q$ is the heat exchanged, $\kappa_i$ is a coefficient related to the thermal conductivity and we have $T_\ell \leq T_\ell^* \leq  T_h^* \leq T_h$. We scale by means of an average temperature $\bar T \equiv (T_h+T_{\ell})/2$ and denote the temperature difference by $\Delta T \equiv T_h-T_{\ell}$. Note that $\theta = 1$ corresponds to   heat conduction as described by Fourier's law. For the transport law \eqref{heatmodified} the calculations are more involved than for \eqref{chemtrans}. For the relative efficiency at maximal power we obtain $e_{mp}/e_c \approx 1/(1+ \theta)$, with a weak dependence on $T_\ell/T_h$, and $e_{mp}/e_c \rightarrow 1$ for $\theta \rightarrow 0$, independently of $T_\ell/T_h$. Once again, ideal efficiency can  be achieved theoretically. This result is consistent with an earlier similar result obtained using a power-law-type heat transport equation \cite{WangTu}. The new result that we add in this context is that the maximum power as a function of $\theta$ is maximal at $\theta = 0$. In the following we provide details of our derivations and show figures to illustrate our main findings. 

\section{Details of the calculations}
\subsection{Chemical}
For the chemical engine the expression for the efficiency $\eta_{mp}$ was obtained in \cite{longpaper}. Here, we evaluate and interpret the maximum power for this engine. We recall
that $\eta$ is the ratio of the mechanical work done by the gas $\Delta N(\mu_h^*-\mu_\ell^*)$ to the chemical work input from the reservoir $\Delta N(\mu_h-\mu_\ell)$ with $\Delta N$ the total number of particles taken up and subsequently released in each cycle.
The power of the engine is obtained by dividing the mechanical work by the total time of the four phases of the cycle \cite{longpaper}. At maximum power the auxiliary chemical potentials take the values, 
\begin{subequations}
\label{eq:mu}
\begin{eqnarray}
\label{eq:mua}
\mu_h^*=\mu_h-\frac{\theta\Delta\mu}{(1+\theta)(1+\gamma)} \\
\label{eq:mub}
\mu_\ell^*=\mu_\ell+\frac{\gamma\theta\Delta\mu}{(1+\theta)(1+\gamma)},
\end{eqnarray}
\end{subequations}
with $\Delta\mu = \mu_h-\mu_\ell$ and $\gamma=(\lambda_h/\lambda_\ell)^\frac{1}{1+\theta}$.

The asymmetry parameter $\gamma$ does not play a crucial role and will typically be assumed to take a value close to unity, in which case we can drop the subscript on the transport coefficient $\lambda$. Possible dependencies of the transport coefficient on geometrical parameters or on the chemical potentials are discussed in \cite{longpaper}. The important parameters are $\Delta\mu$ and $\theta$. Note how the $\mu_i^*$ approach the reservoir values $\mu_i$ when i) the engine operates between two vicinal chemical potentials, i.e., $\Delta \mu \rightarrow 0$, or, ii) when the particle current in the model approaches a step function, i.e., $\theta \rightarrow 0$. It is therefore to be expected that, in the latter limit, ideal efficiency is retrieved. What is surprising, however, is that, in case ii), the engine retains a nonzero power and a nonzero particle current, as we now prove.

Let us first consider the particle current described by \eqref{chemtrans}. When we take the limit $\theta\to 0$ at nonzero $\Delta \mu$ in \eqref{eq:mu} and implement it in \eqref{chemtrans}, we retain a nonzero particle current, since $\lim_{\theta\to 0}|\frac{\mu_i-\mu_i^*}{k_BT}|^\theta=1$. The particle uptake and release times also remain finite in this limit. For arbitrary $\theta$, the uptake and release times, $\tau_1$ and $\tau_3$,  are, respectively, 
\begin{subequations}\label{eq:tau}
\begin{eqnarray}
\label{eq:tau1}
\tau_1=\frac{\Delta N}{\lambda_h k_BT(\frac{\mu_h-\mu_h^*}{k_BT})^\theta}  \\ 
\label{eq:tau3}
\tau_3=\frac{\Delta N}{\lambda_\ell k_BT(\frac{\mu_\ell^*-\mu_{\ell}}{k_BT})^\theta} 
\end{eqnarray}
\end{subequations}

Keeping with the convention that the time required by the other stages of the cycle, isocardinal expansion ($\tau_2$) and compression ($\tau_4$), is just  $\tau_2+\tau_4=(q-1)(\tau_1+\tau_3)$, where $q$ is a (large) constant number \cite{longpaper}, the power of the chemical engine is
\begin{equation}
P=\frac{W}{\sum_{j=1}^{4}\tau_j}
\end{equation}
Using \eqref{eq:mu}, \eqref{eq:tau}, and the result  $\mu_h^*-\mu_{\ell}^*=\Delta\mu/(1+\theta)$, the maximum power of the chemical engine is
\begin{equation}
P_{max}=\frac{\lambda_h k_BT\Delta\mu}{q(1+\theta)(1+\gamma)}\left (\frac{\theta\Delta\mu}{k_BT(1+\theta)(1+\gamma)}\right )^{\theta}
\end{equation}
Taking the limit  $\theta\to 0$, the maximum power becomes 
\begin{equation}
\lim_{\theta\to 0}P_{max}=\frac{\lambda_h k_BT\Delta\mu}{q(1+\gamma)}
\end{equation}

\begin{figure}
\centering
\includegraphics[width=0.44\textwidth]{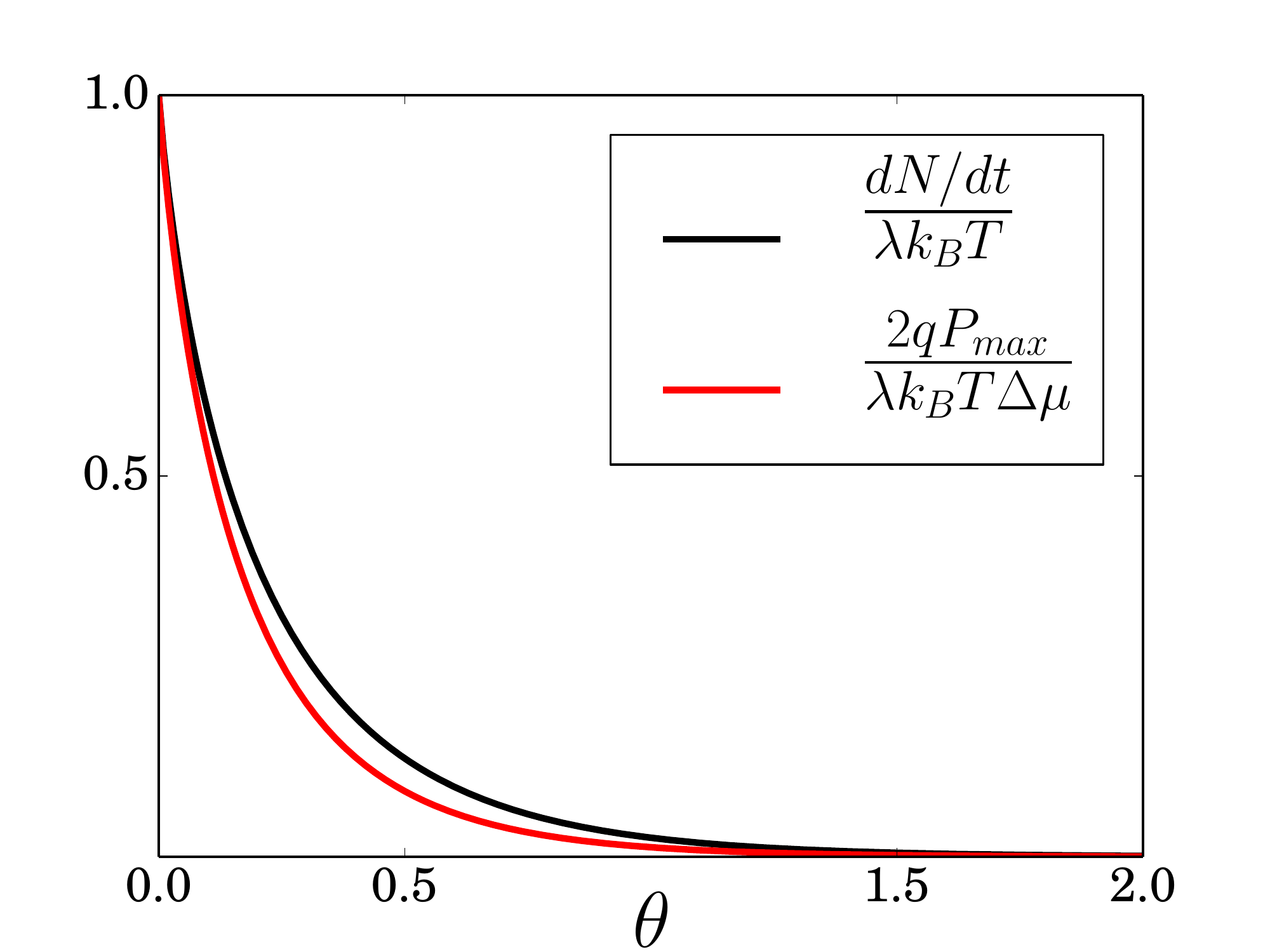}
\caption{Reduced particle current (black; upper curve) and reduced maximum power (red; lower curve) of the chemical engine as a function of the exponent $\theta$ of the transport law. The transport coefficients are assumed equal for uptake and release of particles, \textit{i.e.}, $\gamma=1$ and we have taken $\Delta\mu/k_BT=0.1$.}
\label{fig:1}
\end{figure}

Since in our model the transport coefficient  is assumed to be  independent of the exponent $\theta$ of the transport law, the maximal power for $\theta\to 0$ is not only nonzero but even maximal as a function of $\theta$. This is illustrated in Fig.\ref{fig:1}, which shows the reduced maximum power and corresponding reduced particle current  for a wide range of $\theta$, for $\gamma=1$, and for the case $\Delta\mu/k_B T=0.1$ (an arbitrary value less than of order unity). Note how the maximum power and the particle current are closely related, when considered as a function of the exponent $\theta$ of the transport law. Observe also the important increase in the maximum power for $\theta <1$ as compared to its value for the usual linear law ($\theta = 1$).  

\subsection{Thermal}

The efficiency $e$ of the thermal engine is the ratio of the work done by the fluid $Q_h-Q_{\ell}$ and the heat input $Q_h$. Here, $Q_h=\tau_1(dQ/dt)$ is the heat uptake during isothermal expansion of the gas in a time $\tau_1$, whereas $Q_{\ell}=-\tau_3(dQ/dt)>0$ is the heat rejected during isothermal compression in a  time $\tau_3$. Again, we keep with the convention that the other branches in the cycle, adiabatic expansion (in a time $\tau_2$) and adiabatic compression (in a time $\tau_4$), last just $\tau_2+\tau_4=(q-1)(\tau_1+\tau_3)$, where $q$ is a constant number greater than unity. The power of the thermal engine is
\begin{equation}
P=\frac{Q_h-Q_{\ell}}{\sum_{j=1}^{4}\tau_j}
\end{equation}
Due to the adiabatic processes operating between the auxiliary temperatures $T_h^*$ and $T_{\ell}^*$  we have $Q_h/T_h^*=Q_{\ell}/T_{\ell}^*$, which leads to the following expression for the efficiency of the thermal engine, 
\begin{equation}
e=1-T_{\ell}^*/T_h^*
\end{equation}

\begin{figure}
\centering
\includegraphics[width=0.44\textwidth]{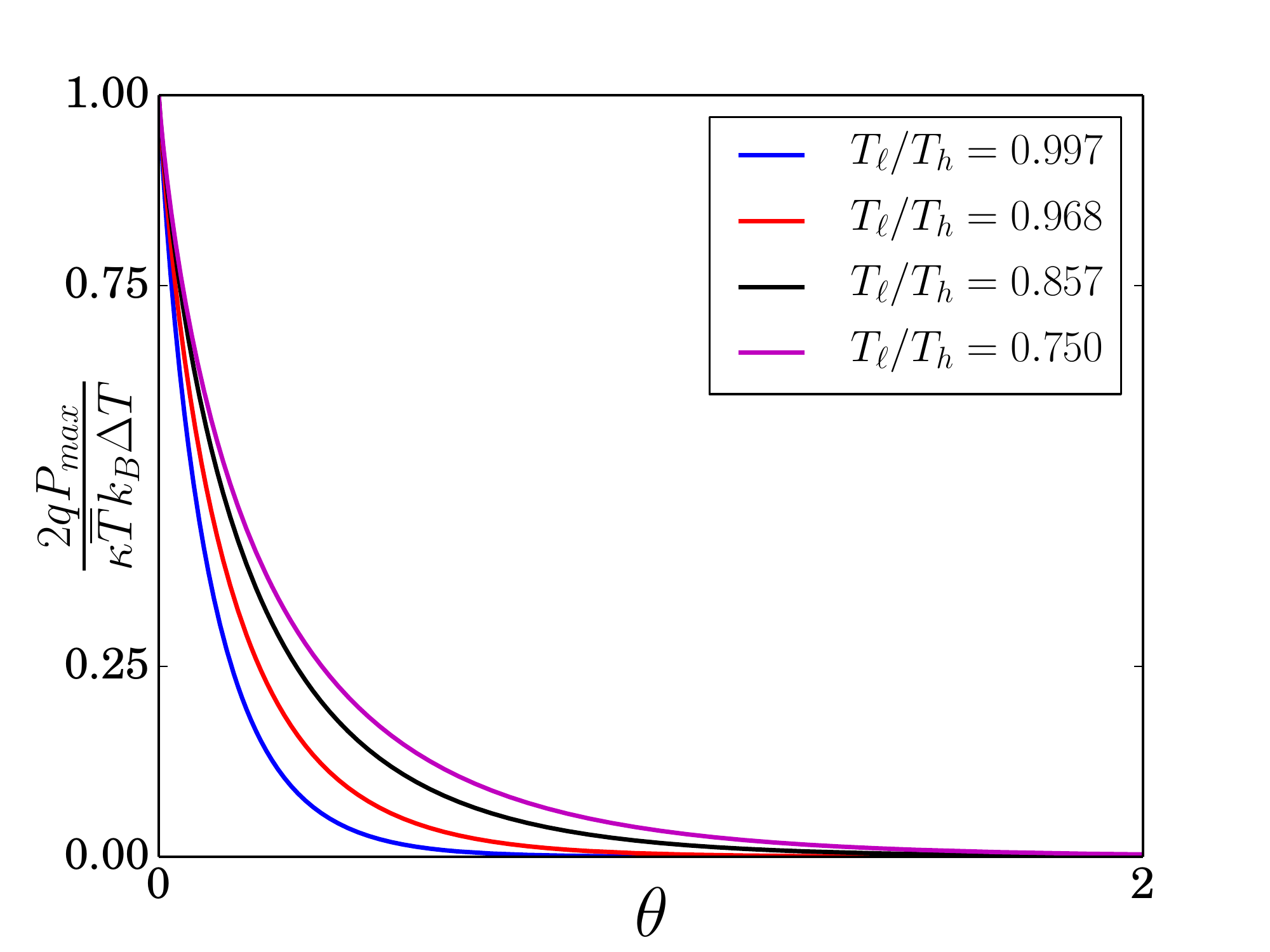}
\caption{Reduced maximum power of the thermal engine as a function of the exponent $\theta$ of the transport law for fixed values of the reservoir temperature ratio $T_{\ell}/T_h$.}
\label{power}
\end{figure}

\begin{figure}
\includegraphics[width=0.44\textwidth]{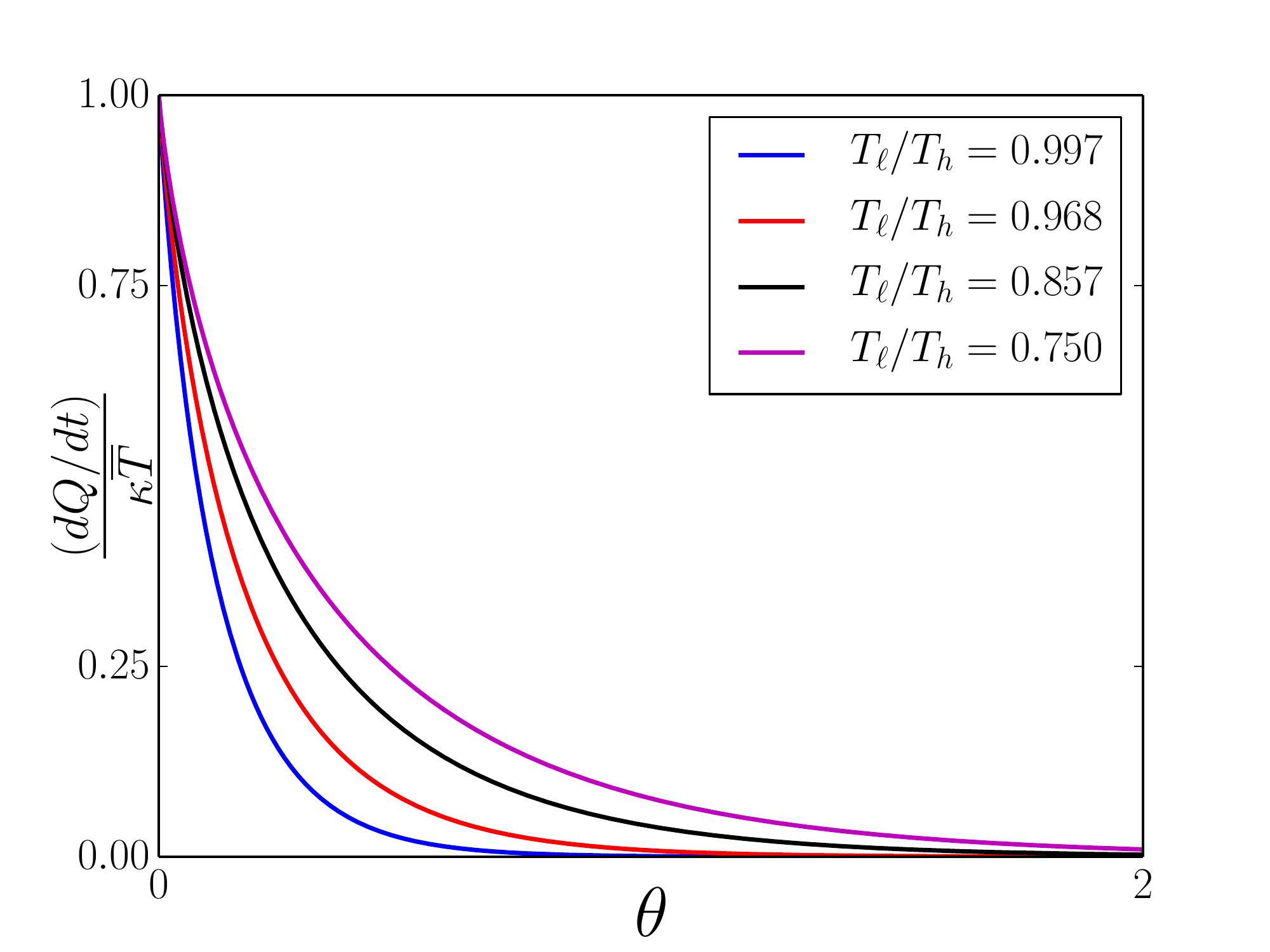}
\caption {The reduced heat input current at maximum power of the thermal engine as a function of the exponent $\theta$ of the transport law for fixed values of the reservoir temperature ratio $T_{\ell}/T_h$.}
\label{heatflux}
\end{figure}

While there is no simple algebraic expression that shows how the auxiliary temperatures -- and therefore the maximum power and the efficiency at maximum power -- depend on $\theta$, a numerical computation is straightforward. We henceforth assume $\kappa_h = \kappa_{\ell} = \kappa$ for simplicity. The result for the maximum power as a function of the exponent $\theta$ of the heat transport law is shown in Fig.\ref{power}. Note how similar is the behavior of the thermal engine to that of the  chemical one (Fig.\ref{fig:1}). For $\theta \rightarrow 0$, the maximum power of the thermal engine is itself maximal. The efficiency at maximum power approaches the Carnot efficiency $e_c$ in this limit. The behavior of the heat input current is shown in Fig.\ref{heatflux}. Also this quantity behaves similarly to its counterpart, the particle current, in the chemical engine (Fig.\ref{fig:1}).

\begin{figure}
\centering
\includegraphics[width=0.44\textwidth]{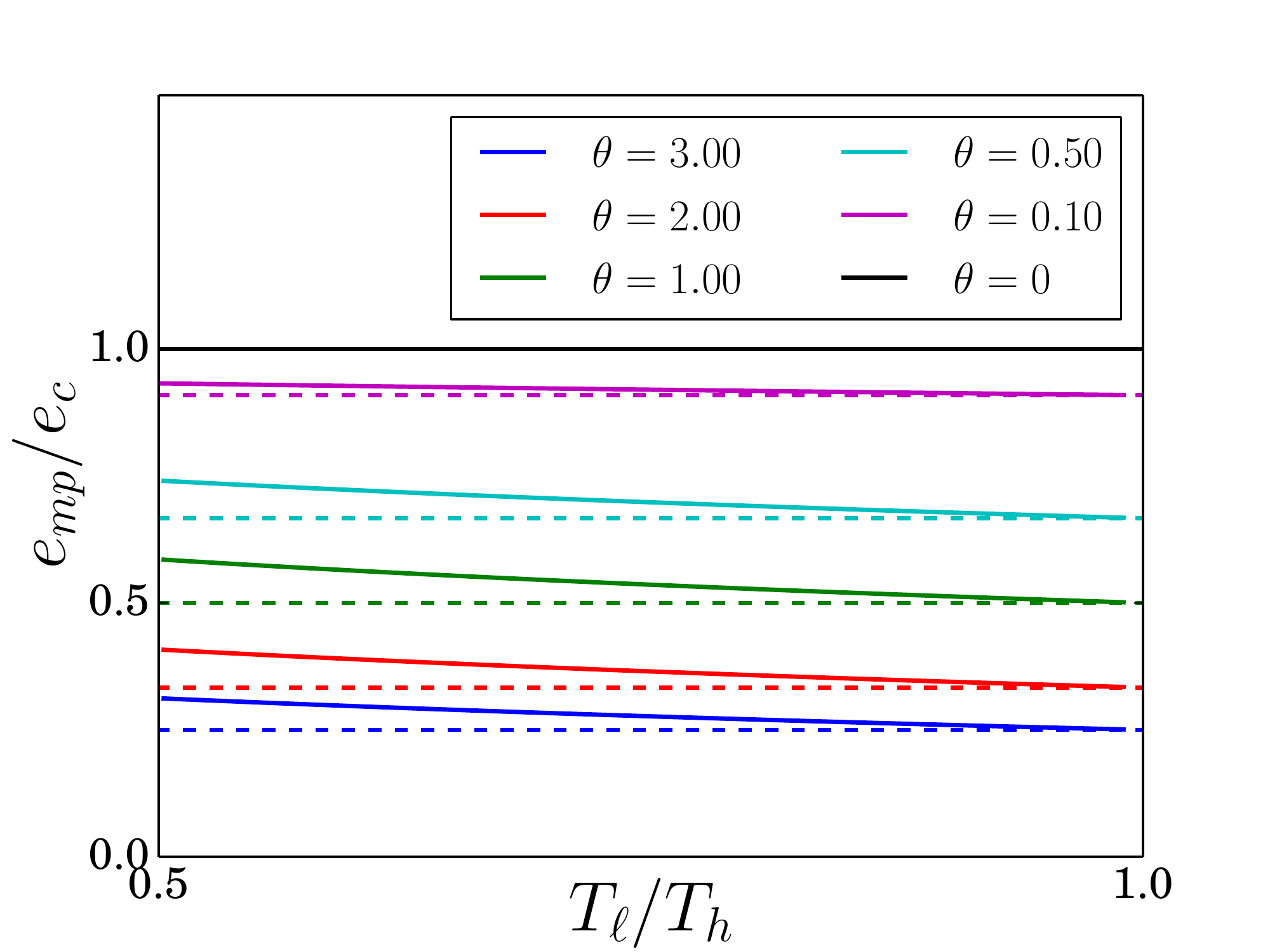}
\caption{Relative efficiencies (solid curves) at maximum power of the thermal engine as a function of temperature ratio of the reservoirs $T_{\ell}/T_h$ for fixed values of the exponent $\theta$ of the transport law. For comparison, the horizontal dashed lines represent the values of the efficiencies at maximum power of the chemical engine, $1/(1+\theta)$.}
\label{fixed_theta}
\end{figure}

\begin{figure}
\centering
\includegraphics[width=0.44\textwidth]{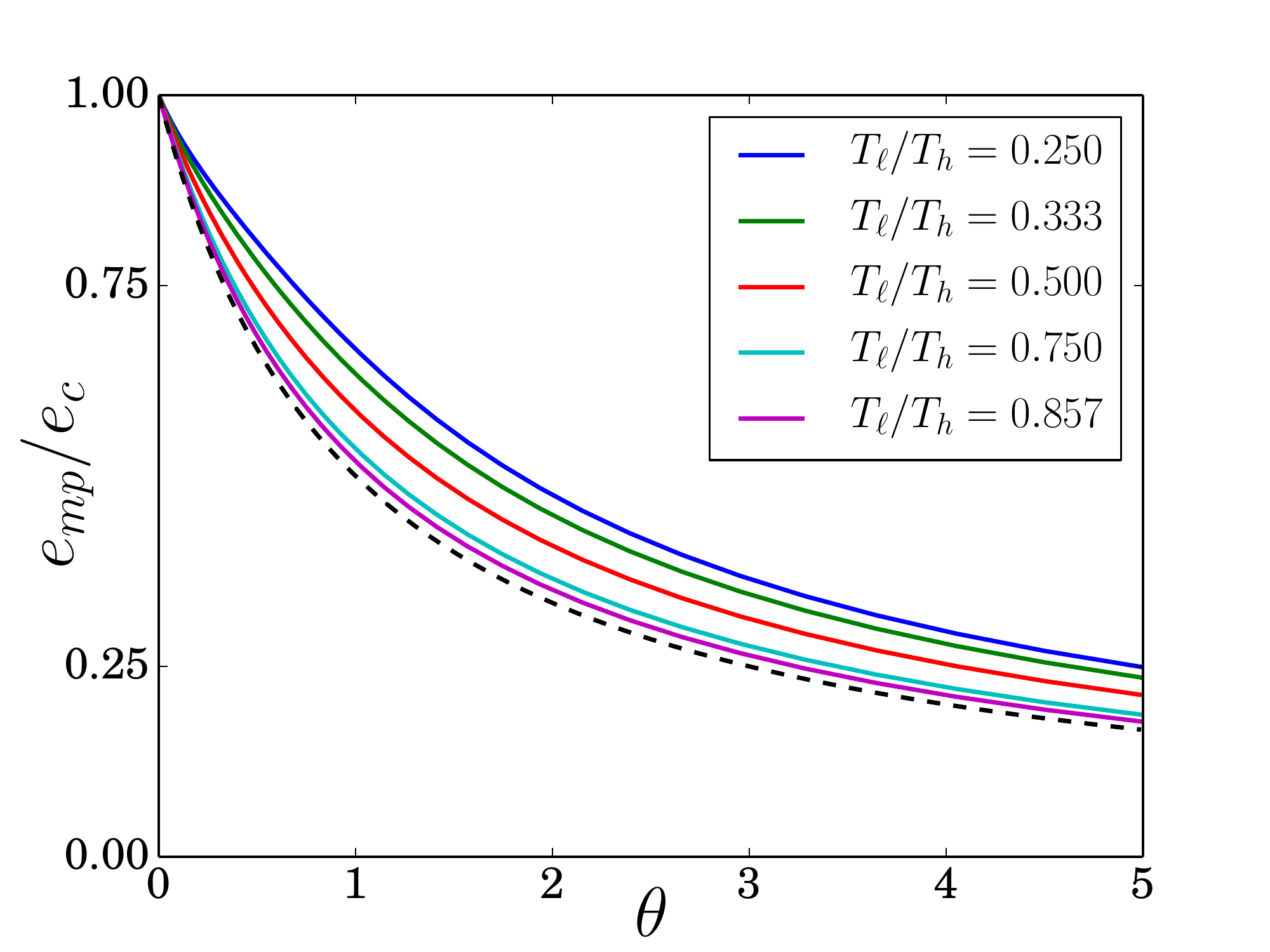}
\caption{Relative efficiencies (solid curves) of the thermal engine as a function of the exponent $\theta$ of the transport law for fixed reservoir temperature ratios $T_{\ell}/T_h$. The dashed curve is the function $1/(1+\theta)$.}
\label{efficiency}
\end{figure}

The differences in properties between the two types of engines are  subtle but  noteworthy. The efficiency ratio $e_{mp}/e_c$ as a function of the reservoir temperature ratio $T_{\ell}/T_h$ is presented for fixed values of $\theta$ in Fig.\ref{fixed_theta}, along with the counterpart efficiencies $\eta_{mp}$ for the chemical engine, which, in contrast, do {\em not} depend on the reservoir chemical potentials. It can be observed that the efficiency ratio $e_{mp}/e_c$ is only weakly dependent on $T_{\ell}/T_h$. The value $e_{mp}/e_c \rightarrow 1/(1+\theta)$ is approached asymptotically from above for $T_{\ell}/T_h \rightarrow 1$. A complementary view is offered in Fig.\ref{efficiency}, which shows the dependence of the efficiency ratio $e_{mp}/e_c$ on $\theta$ for  fixed values of $T_{\ell}/T_h$.

\section{Consistency with thermodynamic inequalities}
As we already mentioned in the introductory section, Shiraishi {\em et al.} \cite{S} derived an exact inequality establishing that a heat engine with nonvanishing power can never attain Carnot efficiency. Their proof assumes any heat engine described by classical mechanics (for $N$ particles, with $N \gg 1$) and for which the interaction with a heat bath can be embodied using the stochastic dynamics of a Markov process. Their main result, restated in our notation, is the inequality
\begin{equation}\label{ineq}
(Q_h + Q_{\ell})^2 \leq \tau \bar \Theta \Delta S,
\end{equation} 
where $\tau \equiv q(\tau_1 + \tau_3) $ is the cycle period and the entropy production $\Delta S$ is given by
\begin{equation}
0 \leq \Delta S = -Q_h/T_h + Q_{\ell} /T_{\ell}
\end{equation}
The left hand side of \eqref{ineq} represents the sum of the absolute values of the heat exchange (squared). If any heat current is present between the system and (any) thermal reservoir, this term is nonzero and positive. The right hand side features the factor $ \bar \Theta $ which depends on the microscopic particle dynamics. In the model employed in \cite{S} this factor is proportional to the (time-averaged) total kinetic energy of the system-reservoir interaction and to the (time-averaged) damping constant determining the coupling between an engine particle and a thermal reservoir of Langevin type.  

The inequality \eqref{ineq} implies the following (weaker) inequality, which is a trade-off relation between power and efficiency \cite{S},
\begin{equation}\label{ineq2}
\frac{W}{\tau} \leq \frac{\bar \Theta}{T_{\ell}} e(e_c-e)
\end{equation} 
From this relationship we infer that the finite power we have obtained in the limit $\theta \downarrow 0$ (step function limit for the current versus applied force) can be consistent with ideal efficiency, $e \rightarrow e_c$, {\em provided} the dynamical factor $\bar \Theta$ diverges in the manner $\bar \Theta \propto 1/(e_c-e) $, or faster. A possible mechanism for this can be a divergence of the damping constant in the Langevin dynamics when the auxiliary temperatures $T_{\ell}^*$ and $T_{h}^*$ approach the reservoir temperatures $T_{\ell}$ and $T_h$, respectively. This, in turn, is plausible in the event of a diverging heat conductivity, since, in linear response theory, the following bound applies \cite{S}
\begin{equation}\label{bound}
\varkappa \lesssim \bar \Theta,
\end{equation} 
where $\varkappa$ is a quantity proportional to the heat conductivity.
Note that, for $\theta =1$ (linear transport law), $\varkappa$ is proportional to the transport coefficient $\kappa$ we have defined through \eqref{heatmodified}. 

In sum, bulk criticality is one of the circumstances in which a diverging heat conductivity can arise. It is then no longer meaningful to assume a linear heat transport equation. Instead, we have proposed to model such a situation using a transport equation with an exponent $\theta < 1$ and a finite coefficient $\kappa$.  In the limit $\theta \downarrow 0$  finite power at ideal efficiency can be achieved without violating the pertinent thermodynamic inequality, provided the divergence of $\bar \Theta$ is allowed for.

\section{Conclusion} 

Identifying possible mechanisms for achieving engines that deliver finite power at ideal efficiency is a nowadays much debated challenge. Several recipes in this regard have been proposed in the context of quantum thermal cycles for small systems. For classical, and macroscopic, systems there have been less proposals. It is widely recognized that working fluids near bulk criticality may significantly enhance the performance of a cycle. Here, we have contributed to these developments by showing that enhanced power can result when the transport law for particle or heat exchange is intrinsically nonlinear, for chemical or thermal cycles, respectively, and that bulk criticality is one relevant circumstance for motivating such a choice of model. 

In foregoing works it was established that the (relative) efficiency at maximum power, which takes the universal value of 1/2 for chemical as well as thermal engines, can in principle be enhanced by considering nonlinear transport laws. The universality is broken only for purely power-law transport laws without linear contributions. Whenever a linear term is present in, or added to, the transport law, the universal value 1/2 for the relative efficiency is recovered in the limit of small $(\mu_h - \mu_{\ell})/k_BT$ or small $(T_h - T_{\ell})/T_h$, but corrections arise for larger values of these differences. Calculations that illustrate these corrections have been presented in \cite{longpaper} for the chemical engine and in \cite{WangTu} for the thermal one.

In this work we have shown that the enhancement of the (relative) efficiency at maximum power can be achieved without giving in on the (particle or heat) current and without losing engine power. Moreover, we have found that, for the simple models considered, the maximum power is itself maximal when the exponent describing the nonlinear transport law is taken to zero, which corresponds to a step-function current. In that limit the (relative) efficiency of the engines approaches unity, which is the ideal value that according to standard thermodynamics considerations can only be reached for a reversible engine. Importantly, we indicate that our result of finite power at vanishing entropy production need not violate the laws of thermodynamics, in particular not the second law. In this respect, we show that for a thermal cycle, our result is consistent with recently proven thermodynamic inequalities, provided the damping constant of the microscopic particle (Langevin) dynamics diverges. Such a divergence can be motivated by a diverging transport coefficient, underscoring the performance-enhancing role of bulk criticality. We call for a more intensive search for possible physical, chemical or biophysical realizations of nonlinear transport models, in view of these surprising properties. 

\section{Acknowledgements}
J.K. and J.O.I. are supported by KU Leuven Research Grant OT/11/063. J.K. thanks Academische Stichting Leuven for support. J.O.I. thanks Jan Naudts for fruitful scepticism and Hans Dierckx for a stimulating question. 

\end{document}